\documentclass[12pt,a4paper]{article}

\usepackage[T1]{fontenc} 
\usepackage{amsmath}
\usepackage{amsfonts}
\usepackage{amssymb}
\usepackage{graphicx}
\usepackage{subfig}
\usepackage{hyperref}
\usepackage{color}
\newcommand{\beq}{\begin{equation}}
\newcommand{\eeq}{\end{equation}}
\newcommand{\bea}{\begin{eqnarray}}
\newcommand{\eea}{\end{eqnarray}}
\def\be{\begin{equation}}
\def\ee{\end{equation}}

\def\beq{\begin{equation}}
\def\eeq{\end{equation}}

\newcommand{\rmd}{\textrm{d}}

\newcommand{\mad}{\mathrm{d}}

\newcommand{\prt}{\partial}

\begin{document}
\def\thefootnote{\fnsymbol{footnote}}

\begin{center}
\Large{\textbf{Khronon inflation}} \\[0.5cm]
 
\large{Paolo Creminelli$^{\rm a}$, Jorge Nore\~na$^{\rm b}$, Manuel Pe\~na$^{\rm c}$, Marko Simonovi\'c$^{\rm d, e}$}
\\[0.5cm]

\small{
\textit{$^{\rm a}$ Abdus Salam International Centre for Theoretical Physics\\ Strada Costiera 11, 34151, Trieste, Italy}}

\vspace{.2cm}

\small{
\textit{$^{\rm b}$ ICC, University of Barcelona (IEEC-UB), Marti i Franques 1, Barcelona 08028, Spain}}

\vspace{.2cm}

\small{
\textit{$^{\rm c}$  Depto. de F\'isica Te\'orica and IFIC, Universdad de Valencia-CSIC, \\ Edificio de Institutos de Paterna, E-46980, Paterna (Valencia), Spain}}

\vspace{.2cm}

\small{
\textit{$^{\rm d}$ SISSA, via Bonomea 265, 34136, Trieste, Italy}}

\vspace{.2cm}

\small{
\textit{$^{\rm e}$ Istituto Nazionale di Fisica Nucleare, Sezione di Trieste, I-34136, Trieste, Italy}}

\vspace{.2cm}

\end{center}

\vspace{.8cm}

\hrule \vspace{0.3cm}
\noindent \small{\textbf{Abstract}\\ 
We study the possibility that the approximate time shift symmetry during inflation is promoted to the full invariance under time reparametrization $t \to \tilde t(t) $, or equivalently under field redefinition of the inflaton $\phi \to \tilde\phi(\phi)$. The symmetry allows only two operators at leading order in derivatives, so that all $n$-point functions of scalar perturbations are fixed in terms of the power spectrum normalization and the speed of sound. During inflation the decaying mode only decays as $1/a$ and this opens up the possibility to violate some of the consistency relations in the squeezed limit, although this violation is suppressed by the (small) breaking of the field reparametrization symmetry. In particular one can get terms in the 3-point function that are only suppressed by $1/k_L$ in the squeezed limit $k_L \to 0$ compared to the local shape.}
\\
\noindent
\hrule
\def\thefootnote{\arabic{footnote}}
\setcounter{footnote}{0}

\section{Introduction}
The approximate scale-invariance of correlation functions produced by inflation is due to the dilation isometry of de Sitter space combined with the approximate symmetry of the inflaton dynamics under time translation \cite{Creminelli:2010ba}
\be
\label{eq:tshift}
 t\rightarrow \tilde t=t+\mathrm{const} \;.
\ee
In this paper we want to explore the possibility that this symmetry is promoted to the full time reparametrization invariance  
\begin{equation}
t\rightarrow \tilde t(t)\;. 
\end{equation}
Of course this symmetry can be a good approximation only during inflation while it must be eventually broken, similarly to what happens with the standard symmetry~\eqref{eq:tshift}, at the end of inflation, when reheating takes place. This symmetry has recently been studied in the context of Ho\v rava gravity and its healthy extensions \cite{Horava:2009uw,Blas:2009qj,DP1}. In these references the scalar mode describing the preferred foliation has been dubbed `khronon'. See \cite{Mukohyama:2009gg,Wang:2009azb, Izumi:2010yn,Huang:2012ep} for other possible connections between Ho\v rava gravity and the creation of primordial curvature perturbations.

We will see that, once this symmetry is enforced, the inflationary dynamics becomes very constrained and unconventional. In particular three features are worth stressing.
\begin{enumerate}
\item All correlation functions of $\zeta$ are fixed, at the lowest order in derivatives, by only two coefficients, which can be written in terms of the normalization of the power spectrum and the speed of sound of perturbations. This is in contrast with the general case, where at any order in perturbations one can write new operators.
\item During inflation the mode wavefunctions have the same form as in Minkowski. This apparently suggests the lack of a proper production of scalar perturbations. However, as we will argue below, this is not true if one considers the inevitable transition to a phase in which the time-reparametrization symmetry is broken.
\item The above feature leaves an interesting signature in the correlation functions of the model. Indeed, the "decaying" mode decays much slower than in the conventional case (as $1/a$ instead of $1/a^3$). This has remarkable consequences for the squeezed limits of correlation functions: the standard single-field theorems hold, but only at first order in the momentum of the long mode. One finds corrections at first order and, in particular, one has a $1/k_L^2$ behaviour of the 3-point function in the squeezed limit. Unfortunately, these effects are very suppressed and totally unobservable. Indeed, the field redefinition symmetry itself is such that a time-dependent background wave, which would violate the consistency relations, can be removed and set to zero. Therefore, these effects are not there in the limit of exact field redefinition symmetry and they will only appear once we consider the small breaking of the symmetry.
\end{enumerate}
Section \ref{sec:action} describes the construction of the action compatible with the $t\rightarrow \tilde t(t)$ symmetry. The power spectrum is studied in Section \ref{sec:power}, with some details left to the two Appendices. The 3- and 4- point functions are discussed respectively in Section \ref{sec:3pf} and \ref{sec:4pf}, while conclusions are drawn in Section \ref{sec:conclusions}.

\section{\label{sec:action}Derivation of the action}

We want to write an inflaton action in which the usual (approximate) symmetry $\phi \to \phi + c$ is promoted to the full invariance under field redefinition $\phi \to \tilde\phi(\phi)$.
We are going to assume an exact de Sitter metric and take the decoupling limit $M_P \to \infty$, in which the dynamics of the scalar perturbations can be studied without considering the mixing with gravity. We will check the validity of this approximation in Appendix \ref{app:constraints}.
The time dependent inflaton background defines a foliation and in the presence of $\phi$ reparameterization invariance, the only invariant object is the 4-vector perpendicular to the foliation \cite{DP1}
\beq
u_\mu=\frac{\prt_\mu \phi}{\sqrt{-g^{\alpha \beta}\prt_\alpha \phi \prt_\beta \phi}}\;, \label{symm}
\eeq
which is indeed invariant under $\phi \to \tilde\phi(\phi)$. Notice that we are requiring a non-zero time-like $\partial_\mu\phi$. At low energy the operators with the smallest number of derivatives will dominate. It is straightforward to realize that it is not possible to write an operator with a single derivative. With two derivatives we have
\beq
\left ( \nabla_\mu u^\mu \right )^2; \quad \nabla_\mu u^\nu\nabla_\nu u^\mu; \quad \nabla_\mu u^\nu\nabla^\mu u_\nu; \quad u^\mu u^\nu\nabla_\mu u_\rho \nabla_\nu u^\rho .
\eeq
The first two are the same by integration by parts (this is true in the de Sitter limit where the Riemann tensor is proportional to the metric). Another constraint comes from the fact that $u^\mu$ is hypersurface-orthogonal, so that the Frobenius theorem implies
\be
 \nabla_\mu u^\nu\nabla_\nu u^\mu = \nabla_\mu u^\nu\nabla^\mu u_\nu + u^\mu u^\nu\nabla_\mu u_\rho \nabla_\nu u^\rho \;.
\ee
We are thus left with two independent operators. The action to lowest order in derivative---and any order in $u^\mu$---can thus be written as
\beq
S=\frac{1}{2}\int\mathrm{d}^4x\sqrt{-g} \left(M_{Pl}^2  R - 2 \Lambda - M_\lambda^2\left(\nabla_\mu u^\mu - 3 H\right)^2+ M_\alpha^2 u^\mu u^\nu\nabla_\mu u_\rho \nabla_\nu u^\rho \right) , \label{S1}
\eeq
where $M_\alpha$ and $M_\lambda$ are the two parameters of our model, besides the vacuum energy $\Lambda$ which is driving inflation. We subtracted $3 H$ from the term proportional to $M_\lambda^2$ to reabsorb its contribution to the vacuum energy in $\Lambda$ (notice that the cross term $\propto \nabla_\mu u^\mu$ is a total derivative). This action gives, at lowest order in derivatives, all the $n$-point functions and it will be the starting point for our calculations below.

Another equivalent way to describe the model is by following the general construction of \cite{Cheung:2007st}. Any inflation model can be described in terms of the metric, in the gauge in which the inflaton perturbations are set to zero. One has to write operators invariant under time-dependent space diffeomorphisms and (approximately) invariant under time translations \cite{Cheung:2007st}
\beq
\label{eq:EFTsymm}
x_i\rightarrow \tilde{x}_i\left({\bf x}, t \right ); \quad t\rightarrow \tilde t=t+\mathrm{const}.
\eeq
Here we promote the symmetry of the inflationary action to \cite{Horava:2009uw,DP1}
 \beq
x_i\rightarrow \tilde x_i\left({\bf x},t\right); \quad t\rightarrow \tilde t(t)\;; \label{Symm}
\eeq
the symmetry $\phi \to \tilde\phi(\phi)$ becomes invariance under time reparametrization, as in this gauge constant time surfaces coincide with the ones at constant inflaton.
Notice that the time reparametrization symmetry forbids to write operators with $g^{00}$, which are otherwise allowed by the symmetries \eqref{eq:EFTsymm}. 
The action \eqref{S1} can be written geometrically as
\beq
S=\frac{M_{P}^2}{2}\int \mad^3 x\,\mad t \sqrt{h} N \left (R^{(3)} +K_{ij}K^{ij}-\lambda (K-3H)^2+\alpha a_ia^i \right ), \label{S_uni}
\eeq
in terms of the ADM variables
\be
ds^2 = -N^2 dt^2 +h_{ij}(d x^i + N^i dt)(d x^j + N^j dt) \;,\qquad K_{ij}=\frac{1}{2N}\left(\dot h_{ij}-\nabla_iN_j-\nabla_jN_i\right)\;,
\label{adm}
\ee
and $a_i\equiv N^{-1}\prt_iN$. 
Indeed in this gauge one has $\left( u^\nu\nabla_\nu u^\mu\right)^2 = a_i a^i$ and $\left(\nabla_\mu u ^\mu\right)^2 = K^2$, so that the equivalence of the two actions follows from the Gauss-Codazzi relation ($R^{(4)} = R^{(3)} +K_{ij}K^{ij}-K^2$ up to total covariant derivatives), with the identification $(\lambda-1) M_P^2 = M_\lambda^2$ and $\alpha M_P^2 = M_\alpha^2$ . Notice that in this language there are four invariant operators with two derivatives: $R^{(3)}$, $K_{ij}K^{ij}$, $ K^2$ and $a_ia^i$. One can get rid of one with the Gauss-Codazzi relation, up to a redefinition of the Planck mass. We still have an additional operator compared to the previous description. Indeed $R^{(3)}$ does not play any role in the decoupling limit. Even more: as it is clear when one changes to spatially flat gauge, where $R^{(3)}$ only depends on tensor modes, this operator does not affect scalar perturbations even departing from the decoupling limit, or at non-linear order. This operator changes the speed of sound of gravitational waves as it affects their spatial kinetic term, but its effect is anyway negligible unless its coefficient is of the order $M_P^2$ (\footnote{Notice also that one cannot induce sizeable graviton non-gaussianities cranking up the coefficient of this operator: indeed its coefficient cannot become parametrically large compared to $M_P^2$, as this would imply a superluminal propagation of tensor modes.}).

The reader may be puzzled by the fact that the symmetry under field redefinition is incompatible with the fact that inflation must end once a certain point in field space is reached. But the situation is not different from the case of the usual shift symmetry, which will be strongly broken at reheating. Also here we only assume the field redefinition symmetry to be a good approximation while inflation occurs and perturbations are generated. Notice that a strong breaking of the symmetry in a region of field space where reheating takes place will not spoil the symmetry somewhere else, as renormalization is local in field space. In other words, the symmetry is valid only in a limited range of field space and it is badly broken if one considers field redefinitions which are large enough to move the point out of the symmetric region

\section{\label{sec:power}Power spectrum}

To calculate the power spectrum we expand the action \eqref{S1} at second order. Using the field redefinition symmetry we can assume to perturb around $\phi_0 = t$, i.e.~$\phi\left({\bf x} , t\right)=t+\pi\left({\bf x},t\right)$, in an unperturbed de Sitter space, which is a good approximation in the decoupling limit. Notice that the action does not contain any term linear in $\pi$, which implies that the unperturbed Universe we are expanding around is indeed a good solution.  In conformal time we get\footnote{The $\pi$ exchange may induce spatial non-locality when coupled to other fields, as discussed in \cite{Blas:2009qj}. This is not relevant for us as we are not interested in coupling with other particles in calculating primordial correlation functions. Spatial non-locality may be relevant in discussing the horizon and flatness problem.}
\begin{align}
S_2 &= \int \mad^3x\,\mad\eta \left(\frac{M_\alpha^2}{2}(\partial\pi')^2 - \frac{M_\lambda^2}{2}(\partial^2\pi)^2 \right) \;.\label{S2}
\end{align}
This result is pretty unconventional. First of all, compared with the usual free-field action, each term has two additional spatial derivatives. This is not worrisome as additional spatial derivatives do not introduce extra pathological degrees of freedom. Second, the action does not contain any $\eta$ dependence so that the field is not sensitive to the expansion of the Universe and behaves as in Minkowski space (though with a speed of sound which is, in general, different from the speed of light). Actually these two peculiarities in some sense cancel each other  to give a scale-invariant spectrum. Indeed, we expect the mode functions to be of the Minkowski form, but with an additional factor of $1/k$ because of the presence of the additional spatial derivatives. It is easy to get the wavefunctions 
\begin{equation}
\label{eq:mode}
\pi_k(\eta)=\frac{1}{\sqrt{2k^3}}\frac{1}{\sqrt{M_\alpha M_\lambda}} e^{\pm i\frac{M_\lambda}{M_\alpha}k\eta} \;,
\end{equation}
which give a scale-invariant spectrum for $\pi$ at late times $\eta \to 0$. The curvature perturbation $\zeta$ is given by $\zeta=-H\pi$ so that 
\begin{equation}
\label{eq:spectrum}
\langle \zeta_{\vec{k}} \zeta_{\vec{k}'} \rangle = (2\pi)^3 \delta (\vec{k}+\vec{k}') \frac{1}{2k^3} \frac{H^2}{M_\alpha M_\lambda}.
\end{equation}
Notice that the scale invariance of the power spectrum (and of higher-order correlation functions) can be justified by symmetry arguments \cite{Creminelli:2010ba}, since we are in exact de Sitter and the action is shift symmetric. Of course, a small tilt is induced if the field redefinition symmetry is slightly broken.

The result is encouraging, but the reader may be suspicious of this derivation. After all, how is it possible that perturbations are created if the field behaves as in Minkowski space? To understand what happens, let us follow the classical dynamics of a given Fourier mode. Although it is not sensitive to the Hubble friction, its wavelength is stretched  and it eventually becomes much longer than the Hubble radius. In this regime the frequency of the mode, which keeps on oscillating as in Minkowski, becomes much slower than the rate of the expansion of the Universe. This means that, on a Hubble timescale, the time-dependence of the mode can be neglected and, similarly, its space-dependence becomes very small in a Hubble patch. We conclude that the solution we are describing is an attractor since the effect of perturbations becomes smaller and smaller as time evolves. 

This also sheds light on the quantum mechanical behaviour. Although each Fourier mode effectively remains in Minkowski, hindering a classical interpretation, the fact that its frequency becomes much smaller than the rate of expansion means that one is sensitive only to $\pi$ and not to $\dot\pi$. It is like probing in a laboratory a harmonic oscillator with an experiment which is very short compared to the period of oscillation: it will only be sensitive to the probability distribution of the position, but not to the momentum. The difference with the standard situation in inflation is quantitative, but not qualitative. Usually the time dependence of the mode decays, compared with the Hubble rate, as $a^{-3}$ and it can safely be neglected. Here it decays as $a^{-1}$. 

The same logic also implies another important result: the conservation of $\zeta$ on super-Hubble scales during the reheating stage and later. Independently of the details of reheating, we can assume that it will be insensitive to $\dot\pi$ which is exponentially small compared to $\pi$. This means that locally we are following the same unperturbed solution, with $\zeta$ describing the relative difference in expansion between different points. In Appendix \ref{app:after} we verify these intuitive arguments in an explicit toy example. We will see, in the following Sections, that this slow decay of the decaying mode leaves some signature in the higher-order correlation functions, which is a quite distinctive feature of this model.

Due to the field redefinition symmetry one can choose  the background solution to be $\phi_0 = - \eta$ and perturb now around this background $\phi = - \eta + \chi$. It is straightforward to express at linear order these perturbations in terms of the perturbations around cosmic time as $\chi = \pi / a$, and write the second order action in terms of $\chi$ from equation \eqref{S2}
\begin{equation}
S_2(\chi) = \int \mathrm{d}^3x\,\mathrm{d}\eta\,a^2 \left(\frac{M_\alpha^2}{2}(\partial\chi')^2 - \frac{M_\lambda^2}{2}(\partial^2\chi)^2 - M_\alpha^2 \mathcal{H}^2 (\partial\chi)^2 \right) \;.
\end{equation}
This is compatible with the results of \cite{Blas:2011en}, where it was noted that the effective mass is that of a conformally coupled field; this is consistent with the fact that the equations of motion for the field are like in Minkowski. Moreover, note that this action gives a power spectrum for $\chi$ which is still scale invariant (since $\chi$ and $\pi$ are related simply by a function of time) but with an amplitude that decreases exponentially during inflation.  Different choices for the background solution seem to give different answers for the power spectrum in spite of the field redefinition symmetry. The issue is settled by the fact that what is more closely related to observations is the curvature perturbation conserved outside of the horizon $\zeta$ which is equal to $\pi$ up to a constant factor as computed in Appendix \ref{app:constraints}.

\section{\label{sec:3pf}The 3-point function}

As we saw in the previous Section, the power spectrum for the fluctuations is scale invariant and indistinguishable from the predictions of more conventional inflationary scenarios. Let us now study the 3-point correlation function which carries additional information. It is conventional to define
\begin{equation}
\langle\zeta_{\vec{k}_1}\zeta_{\vec{k}_2}\zeta_{\vec{k}_3}\rangle \equiv (2\pi)^3\delta(\vec{k}_1+\vec{k}_2+\vec{k}_3) F_\zeta(k_1, k_2, k_3)\,,
\end{equation}
where translational invariance implies that the 3-point function must be proportional to the Dirac delta, and rotational invariance implies that the function $F_\zeta$, called the bispectrum of $\zeta$, is a function only of the magnitude of the momenta. As discussed in the previous Section, the dilation isometry of de Sitter, together with the time shift symmetry implies that the bispectrum is a homogeneous function of degree $-6$.

The 3-point function of the field perturbation $\pi$ can be computed using the in-in formalism.  It is given by (see Ref.~\cite{Maldacena:2002vr})
\begin{equation}
\langle\pi^3(\eta_\ast)\rangle = \bigg\langle 0\bigg| \bar Te^{i\int_{-\infty-i \epsilon}^{\eta_\ast} H_{\mathrm int}(\eta')\mathrm{d}\eta'} \pi^3(\eta_\ast)Te^{-i\int_{-\infty + i \epsilon}^{\eta_\ast} H_{\mathrm int}(\eta')\mathrm{d}\eta'} \bigg| 0\bigg\rangle\,,
\end{equation}
where $|0\rangle$ is the Bunch-Davies vacuum, $T$ ( and $\bar T$) indicates time ordering (and anti time ordering), $\eta_\ast$ indicates the time at which inflation ends, $\mathcal{H}_{\mathrm{int}}$ is the interaction Hamiltonian and $\epsilon$ is an infinitesimal positive constant. At leading order only the cubic part of the interaction Hamiltionian contributes, and one can show that $\mathcal{H}_{\mathrm{int}} = - \mathcal{L}_{\mathrm{int}}$. Therefore one can use the third order piece of the Lagrangian to compute the three-point function using
\begin{equation}
\langle\pi^3(\eta_\ast)\rangle = i \int_{-\infty}^{\eta_\ast} \mathrm{d}\eta\,\bigg\langle \bigg[\pi^3(\eta_\ast),\int\mathrm{d}^3x\,\mathcal{L}_{\mathrm{int}}(t,\vec{x})\bigg]\bigg\rangle\,.
\label{in-in, 3pf}
\end{equation}
The interaction Lagrangian can be computed by expanding the action, equation \eqref{S1}, to third order.  We get, after several integrations by parts\footnote{The reader might be worried about the appearance of an interacting term in the action that contains explicitly a second time derivative acting on $\pi$ that cannot be removed by a partial integration. This would not be a problem for us as we are treating these higher derivative terms as small corrections to the free action. 
However, it was noted in \cite{DP1} that this term can actually be reabsorbed by performing a field redefinition of the form
\beq
\pi=\bar\pi+\bar\pi\bar\pi', \label{f_redef1}
\eeq
leading to the following action
\begin{equation}
S_3 = \int \mad^3x\,\mad\eta\,\frac{1}{a} \Big[- M_\lambda^2 \Big(\bar\pi \prt^2\bar\pi \prt^2\bar\pi'  -\frac{\mathcal{H}}{2}(\prt\bar\pi)^2\prt^2\bar\pi \Big) + M_\alpha^2  \Big(\frac{1}{2}\bar\pi(\prt\bar\pi')^2-\frac{\mathcal{H}}{2}\bar\pi(\prt\bar\pi')^2 - \partial_i\bar\pi'\partial_j\bar\pi\partial_i\partial_j\bar\pi \Big)\Big]\,.
\label{S3}
\end{equation}
This action produces the same three-point function, eq. \eqref{3pf}, since the field redefinition \eqref{f_redef1} vanishes outside of the horizon. One expects this to be true at every order in perturbations since in the unitary gauge, eq. \eqref{S_uni}, the number of degrees of freedom is fixed \cite{DP1}.
},
\begin{equation}
S_3 = \int \mathrm{d}^3x\,\mathrm{d}\eta\, \frac{1}{a}\Big[M_\lambda^2 \big(2\partial_i\pi'\partial_i\pi\partial^2\pi + \pi' \partial_i\partial_j\pi\partial_i\partial_j \pi \big) + M_\alpha^2 \big(\pi'\partial_i\pi''\partial_i\pi - \partial_i\pi'\partial_j\pi\partial_i\partial_j\pi \big)\Big]\,.
\label{S3old}
\end{equation}
This cubic action coincides with eq.~(5.10) of \cite{DP1} in the Minkowski limit. In order to compute the 3-point function for $\zeta$ we use the relation $\zeta = -H\pi$, additional non-linear terms in this relation either involve higher derivatives, which vanish outside of the horizon, or are suppressed by slow-roll factors, see
 \cite{Maldacena:2002vr,Cheung:2007sv}. We thus obtain the following expression for the bispectrum:
\begin{multline}
F_\zeta(k_1,k_2,k_3) = \frac{1}{\prod k_i^3} P_\zeta^2\bigg[-\frac{k_1}{k_t^2}(k_3^2\vec{k}_1\cdot\vec{k}_2 + k_2^2\vec{k}_1\cdot\vec{k}_3) - \frac{k_1^2}{k_t}\vec{k}_2\cdot\vec{k}_3 - \frac{M_\alpha^2}{M_\lambda^2}\frac{k_1^3}{k_t^2}\vec{k}_2\cdot\vec{k}_3\bigg] \\ +\, \mbox{cyclic perms.}\,,
 \label{3pf}
\end{multline}
where $k_t \equiv k_1 + k_2 + k_3$ and $P_\zeta = H^2/(2 M_\alpha M_\lambda)$ is the $\zeta$ power spectrum, eq.~\eqref{eq:spectrum}. All the contributions but the last cannot be large and give an $f_{\rm NL} \sim 1$. The contribution from the last term on the other hand is proportional to $M_\alpha^2/M_\lambda^2 \equiv 1/c_s^2$. Actually it is easy to estimate the effect of each operator of the cubic action \eqref{S3old} comparing them with the quadratic action when modes freeze ($\partial_t \sim H$, $\partial_i \sim H/c_s$). The only operator that can give a parametrically large 3-point function is the last in eq.~\eqref{3pf}.

\begin{figure}[!htb]
\begin{center}
\includegraphics[width=0.6\textwidth]{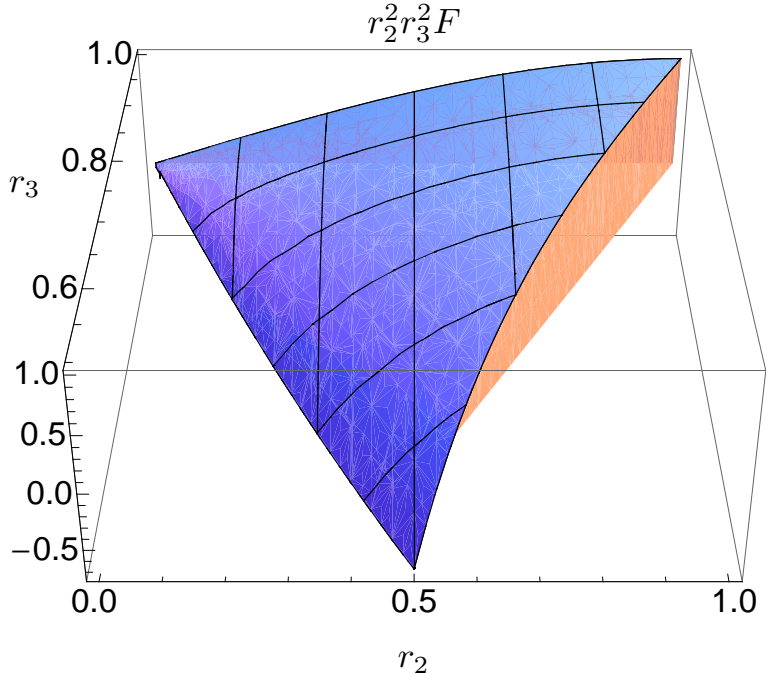}
\end{center}
\caption{\small {\em We plot the shape of the part proportional to $1/c_s^2$ of the 3-point function, equation \eqref{3pf}, as a function of the ratios between momenta $r_2 \equiv k_2/k_1$ and $r_3 \equiv k_3/k_1$, multiplied by $r_2^2 r_3^2$. The shape is normalized such that its amplitude is one at the equilateral point $r_2 = r_3 = 1$.}}
\label{shape}
\end{figure}

We find an interesting feature of the model: it gives a single potentially large shape with an amplitude controlled by a single parameter, namely $c_s^2$.  We plot the shape of this contribution in figure \ref{shape}. 

In order to understand the phenomenological implications of this result, let us first introduce a quantitative way of comparing bispectra. One defines the scalar product between two shapes as \cite{Babich:2004gb}
\begin{equation}
F_1 \cdot F_2 = \sum_{\mathrm{triangles}} \frac{F_1(k_1,k_2,k_3) F_2(k_1,k_2,k_3)}{P(k_1)P(k_2)P(k_3)}\,,
\end{equation}
where the sum is over values of the momenta that form a closed triangle. One can then define the ``cosine'' of two shapes in the following way
\begin{equation}
\cos(F_1,F_2) = \frac{F_1\cdot F_2}{(F_1\cdot F_1\,F_2\cdot F_2)^{1/2}}\,.
\end{equation}
If the cosine between two shapes is close to one, one expects the data to be unable to distinguish between the two; conversely, if the cosine between two shapes is very small, constraints on the amplitude of one of the shapes do not constrain the amplitude of the other\footnote{For CMB applications, this statement can be made more precise by defining a ``two-dimensional'' cosine, which takes into account the geometry and the effect of the linear transfer functions, to get closer to what it is actually observed \cite{Babich:2004gb}.}. 

In CMB data analysis, a crucial numerical boost is gained when looking for shapes which are factorizable, \emph{i.e.}~which can be written as monomials of $k_1$, $k_2$, and $k_3$. The standard procedure when comparing a theoretical 3-point function with constraints from CMB data is then to look for a factorizable shape which has a large cosine with the shape generated by a given inflationary model. Such shapes are often termed templates, which can be expressed as linear combinations of the so-called local, equilateral and orthogonal templates (see refs.~\cite{Babich:2004gb,Senatore:2009gt}). The cosines of the shape depicted in figure \ref{shape} with these three standard templates are
\begin{align}
\cos(F_\zeta, F_{\mathrm{local}}) &= 0.17\,,\\
\cos(F_\zeta, F_{\mathrm{equilateral}}) &= 0.93\,,\\
\cos(F_\zeta, F_{\mathrm{orthogonal}}) &= 0.49\,.
\end{align}
It is therefore a good approximation to take the shape as equilateral. Its amplitude can be read from the expression \eqref{3pf} above
\be
f_{\rm NL}^{\rm eq} = \frac{5}{108} \frac{1}{c_s^2} \;.
\ee
The limits on the equilateral shape obtained from WMAP 7 data given in ref.~\cite{Komatsu:2010fb} can be used to put bounds  on $c_s$:  $c_s \gtrsim 0.013$ at the $95\%$ confidence level. Notice that (the potentially large contribution to) the 3-point function has a fixed positive sign in this model\footnote{We are using the WMAP sign convention for $f_{\rm NL}$.}. This is the opposite of what happens in more conventional models with reduced speed of sound ($K$-inflation), where the operator which reduces the speed of sound gives $f_{\rm NL}^{\rm eq} \propto -1/c_s^2$. However in those models one has another operator which contributes to the 3-point function and can flip the sign of $f_{\rm NL}^{\rm eq}$; in our case we have no freedom. It is worth stressing that, although the shape given in eq.~\eqref{3pf} has a large overlap with the equilateral one, the result has no free parameter and thus represents a potential smoking gun of the model.

If one calculates the contribution to the 3-point function of the second and third operator in the action \eqref{S3old}, each of them, when taking the squeezed limit $k_1 \ll k_2, k_3$, diverges like $1/k_1^2$ (while the leading term discussed above goes as $1/k_1$). This seems to contradict the results of references \cite{Creminelli:2011rh,Creminelli:2012ed} where it is shown that, ignoring small deviations from scale invariance, in single field inflationary models the squeezed limit of the 3-point function diverges like $1/k_1$ in that squeezed limit. This is due to the fact that the proof given there relies on $\zeta'_k(\eta)$ vanishing at least like $k^2/H^2$ outside of the horizon, while equation \eqref{eq:mode} shows that in the model we are studying here $\zeta'_k(\eta)$ vanishes only like $k/H$. However, this $1/k_1^2$ divergence cancels between the two operators as it is evident in \eqref{3pf} (\footnote{We are indebted to Austin Joyce for pointing out an error in the first version of the paper.}). This is not an accident, but a consequence of the field redefinition symmetry. Indeed, a homogeneous time-dependent background mode, which would lead to the violation of the consistency relations, can be redefined away using the symmetry.
Our theory is invariant under
\be
\label{eq:redsym}
t +\pi \to F(t + \pi) \simeq  t + \pi +\epsilon(t+ \pi) + \ldots = t + \pi + \epsilon(t) + \dot\epsilon(t) \pi + \frac12 \ddot\epsilon(t) \pi^2  + \ldots
\ee
This means that a time-dependent background $\epsilon(t)$ can be removed, provided we also redefine the field $\pi$ as above. This redefinition, however, is irrelevant at late times as $\dot\epsilon \to 0$, so we can conclude that the background wave has no effect and thus we do not violate any consistency relation. We checked explicitly that terms obtained from the cubic action with $\pi \to \pi + \epsilon(t)$ cancel with terms in the quadratic action \eqref{S2} that appear after $\pi \to \pi + \dot\epsilon(t) \pi$. 

Notice, however, that the slow decay of the modes still opens up the possibility to violate the consistency relations: it is enough to consider terms which violate the original field redefinition symmetry. Of course, we expect these terms to be suppressed but to be there anyway, as indicated by the small observed deviation from a scale invariant spectrum. For example, we can add to the action \eqref{S_uni} the cubic operator in unitary gauge
\be
S \supset \int \mad^3 x\,\mad t \sqrt{h} N \;(g^{00}-1)(K-3H)^2 \;,
\ee
which is {\em not} invariant under the field redefinition symmetry.
This operator starts at cubic order, so that it does not modify the mode evolution, and it gives a cubic term $\pi' (\partial^2\pi)^2$ which will violate the consistency relations.  One might hope that such a non-standard behavior leaves an observational signature for example in the scale-dependence of the halo bias \cite{Dalal:2007cu,Matarrese:2008nc,Afshordi:2008ru}. However, even under the optimistic assumption that the effect is suppressed by a single power of slow-roll compared to the leading $1/c_s^2$ term, the analysis of references \cite{Norena:2012yi,Sefusatti:2012ye} (though performed for a different model) indicates that the observation of this effect with such a small amplitude seems unfeasible with planned surveys.

\section{\label{sec:4pf}The 4-point function}
Given that all the correlation functions to leading order in derivatives are completely fixed by two coefficients, it is of some interest to look at the 4-point function. 
In this section we compute the 4-point function focusing only on the leading contribution proportional to $c_s^{-4}$. This part of the 4-point function is important since observationally it gives the most relevant contribution in the case of small $c_s$. 

In order to compute the 4-point function we need the interaction Hamiltonian to fourth order, for which it is no longer true that $\mathcal{H}_{int} = - \mathcal{L}_{int}$. Let us start by expanding the action \eqref{S1} to fourth order 
\begin{align}
S^{(4)}_\alpha& =M_\alpha^2\int \rmd^3x\,\rmd\eta \left(\frac{H^2}{2} \pi'\pi'(\partial\pi)^2 -\frac{H}{a}\left( (\partial \pi)^2\partial_i\pi\partial_i\pi' + \pi'\pi''(\partial\pi)^2 +\pi'\partial_i\pi\partial_j\pi\partial_i\partial_j\pi \right) \right. \nonumber \\
& \left. - \frac{3H}{a} \pi'\pi'\partial_i\pi\partial_i\pi' + \frac{1}{2a^2} \left( \pi''\pi''(\partial\pi)^2 + 6\pi'\pi'' \partial_i\pi\partial_i\pi' + 3 \pi'\pi' \partial_i\pi'\partial_i\pi'  + 3\partial_i\pi\partial_j\pi\partial_i\pi'\partial_j\pi' \right)\right. \nonumber \\
& \left. + \frac{1}{a^2} \pi''\partial_i\pi\partial_j\pi\partial_i\partial_j\pi + \frac{3}{a^2} \pi'\partial_i\pi\partial_j\pi'\partial_i\partial_j\pi + \frac{1}{a^2}(\partial\pi)^2(\partial\pi')^2 + \frac{1}{2a^2} \partial_i\pi\partial_j\pi \partial_i\partial_l\pi\partial_j\partial_l\pi \right) \;.
\label{S4}
\end{align}
\begin{align}
S_\lambda^{(4)} & =-M_\lambda^2\int  \rmd^3x\,\rmd\eta  \left( \frac{H^2}{8}(\partial \pi)^2 (\partial \pi)^2 + \frac{H}{a} (\partial \pi)^2 \partial_i\pi\partial_i \pi' + \frac{3H}{2a}(\partial \pi)^2 \pi'\Delta\pi \right. \nonumber \\ & \left. + \frac{1}{a^2} \Delta\pi\partial_i\pi\partial_j\pi\partial_i\partial_j\pi + \frac{1}{2a^2} (\Delta\pi)^2(\partial\pi)^2 + \frac{6}{a^2} \pi'\Delta\pi\partial_i\pi\partial_i \pi' + \frac{3}{2a^2} \pi' \pi'\Delta\pi\Delta\pi \right. \nonumber \\
& \left.  + \frac{1}{a^2} \pi''\Delta\pi(\partial\pi)^2 - \frac{H}{a} \pi' \Delta\pi(\partial\pi)^2+ \frac{2}{a^2}\partial_i\pi\partial_j\pi\partial_i \pi'\partial_j \pi' \right) \;.
\label{S42}
\end{align}

The second and third order pieces of the action are given by equations \eqref{S2} and \eqref{S3old} respectively. Throughout this section we will only keep those terms that give the largest contributions to the 4-point function in the small $c_s^2$ case. As before, it is easy to estimate the amplitude of the 4-point function by comparing each term in the quartic action \eqref{S4} and \eqref{S42} with the kinetic terms \eqref{S2} once the modes freeze $(\partial_t \sim H$ and $\partial_i \sim H/c_s$). The amplitude of the largest piece of the 4-point function can thus be estimated to be proportional to $c_s^{-4}$, generated by those terms in the non-linear action which are proportional to $M_\alpha^2$ and containing the highest number of spatial derivatives. Thus, we will keep only the last terms in equations \eqref{S3old} and \eqref{S4}.

As stressed above, in order to obtain the correct expression for the 4-point function one must explicitly compute the Hamiltonian\footnote{Notice that the canonical variables satisfying the commutation relations after quantization are the field $\pi$ and the generalized momentum $P$, and the Hamiltonian is a function of these variables. Wherever we write $\pi'$ in the explicit expression for the Hamiltonian, it should be understood as shorthand for the appropriate expression in terms of $\pi$ and $P$.} $\mathcal{H}(P,\pi)=P\pi' - \mathcal{L}(P,\pi)$, where the generalized momentum (keeping only the most relevant pieces in the small $c_s^2$ case) is given by
\be
P = \frac{\partial\mathcal{L}}{\partial \pi'} = -M_\alpha^2 \partial^2\pi' +\frac{M_\alpha^2}{2a}\partial^2(\partial_i\pi)^2 \;.
\label{P}
\ee
A straightforward computation of the terms in the fourth order interaction Hamiltonian which could potentially generate a 4-point function proportional to $c_s^{-4}$ using equations \eqref{S4} and \eqref{P} shows that it vanishes\footnote{It is important to note that in the full fourth order action, equation \eqref{S4}, there are terms containing two time derivatives acting on the field $\pi''$ in such a way that they cannot be eliminated by an integration by parts. Similarly to what we did in the case of the 3-point function, we could have removed these terms from the action by a suitable field redefinition
\begin{equation*}
\pi=\bar\pi+\bar\pi\bar\pi' + \partial^{-2}\partial_i\left( \frac{1}{a}\bar\pi'\partial_i \bar\pi \right)\;,
\end{equation*}
which vanishes outside the horizon and does not change the result for the correlation functions.
}.

In principle, two types of diagrams can contribute to the 4-point function: exchange diagrams and 
contact diagrams. However, since the fourth order interaction Hamiltonian vanishes, there is no contact diagram and the vacuum expectation value for the four-point equal-time correlation function in momentum space is given only by the exchange diagrams. In the in-in formalism the 4-point function can be then computed as
\begin{multline}
\langle 0 | \zeta_{\vec k_1} \zeta_{\vec k_2} \zeta_{\vec k_3} \zeta_{\vec k_4} (\eta) | 0 \rangle = 
 \int_{-\infty}^\eta \mathrm{d}\eta' \int_{-\infty}^{\eta} \mathrm{d}\eta'' \langle 0 | \mathcal{H}_{int}^{(3)}(\eta') \zeta_{\vec k_1} \zeta_{\vec k_2} \zeta_{\vec k_3} \zeta_{\vec k_4} (\eta)  \mathcal{H}_{int}^{(3)}(\eta'')| 0 \rangle \\
 - 2\;\mathrm{Re} \left( \int_{-\infty}^\eta \mathrm{d}\eta' \int_{-\infty}^{\eta'} \mathrm{d}\eta'' \langle 0 | \zeta_{\vec k_1} \zeta_{\vec k_2} \zeta_{\vec k_3} \zeta_{\vec k_4} (\eta)  \mathcal{H}_{int}^{(3)}(\eta') \mathcal{H}_{int}^{(3)}(\eta'')| 0 \rangle \right)\;, \label{inin}
\end{multline}
The third order interaction Hamiltonian can be read from equation \eqref{S3old}. We are interested in the piece that can give a contribution to the 4-point function proportional to $1/c_s^4$ which, after an integration by parts, we write as
\begin{equation}
\mathcal{H}^{(3)}_{int} = -\frac{M_\alpha^2}{2a}\partial^2\pi'(\partial\pi)^2\;.
\label{H3}
\end{equation}
The time integrations appearing in equation \eqref{inin} can be performed using
\begin{equation}
\int_{-\infty}^0 \frac{\mathrm{d}\tau'}{a(\tau')} e^{-i\frac{M_\lambda}{M_\alpha}(p+k_1+k_2)\tau'}  \int_{-\infty}^0 \frac{\mathrm{d}\tau''}{a(\tau'')} e^{i\frac{M_\lambda}{M_\alpha}(p+k_3+k_4)\tau''} = 
 H^2\frac{M_\alpha^4}{M_\lambda^4}\frac{1}{2p^3}\frac{1}{(p+k_1+k_2)^2} \frac{1}{(p+k_3+k_4)^2}\;,
\label{4pf_integral}
\end{equation}
and
\begin{multline}
\int_{-\infty}^0 \frac{\mathrm{d}\tau'}{a(\tau')} e^{i\frac{M_\lambda}{M_\alpha}(k_1+k_2 -p)\tau'}  \int_{-\infty}^{\tau'} \frac{\mathrm{d}\tau''}{a(\tau'')} e^{i\frac{M_\lambda}{M_\alpha}(p+k_3+k_4)\tau''} =  \\
H^2\frac{M_\alpha^4}{M_\lambda^4}\frac{1}{2p^3} \frac{1}{(p+k_3+k_4)^2} \left( \frac{1}{k_t^2} + 2 \frac{p+k_3+k_4}{k_t^3} \right) \;.
\label{4pf_integral2}
\end{multline}
The 4-point function can then be computed using equations \eqref{inin} to \eqref{4pf_integral2}
\begin{multline}
\langle \zeta_{\vec k_1} \zeta_{\vec k_2} \zeta_{\vec k_3} \zeta_{\vec k_4} \rangle_c = (2\pi)^3\delta(\sum\vec k_a) P_\zeta^3 \frac{M_\alpha^4}{M_\lambda^4} \frac{1}{\prod_a k_a^3}\frac{1}{4 p^3(p + k_1 + k_2)^2} \\
\times\Bigg\{\big(p^6(\vec{k}_1\cdot\vec{k}_2)(\vec{k}_3\cdot\vec{k}_4) - 2 p^3k_1^3(\vec{p}\cdot\vec{k}_2)(\vec{k}_3\cdot\vec{k}_4) \big)\bigg[\frac{1}{4(p+k_3+k_4)^2}-\frac{1}{2k_t^3}\big(k_t + 2(p+k_1+k_2)\big)\bigg] \\
+\big( 2 p^3k_3^3(\vec{k}_1\cdot\vec{k}_2)(\vec{p}\cdot\vec{k}_4) - 4k_1^3 k_3^3(\vec{p}\cdot\vec{k}_2)(\vec{p}\cdot\vec{k}_4) \big)\bigg[\frac{1}{4(p+k_3+k_4)^2}+\frac{1}{2k_t^3}\big(k_t + 2(p+k_1+k_2)\big)\bigg] \Bigg\} \\+ \mbox{23 perms.}\;,
\label{4pf}
\end{multline}
where $\vec{p} = \vec{k}_1 + \vec{k}_2$.

Eq.~\eqref{4pf} is suppressed in the squeezed limit and does not contribute to the consistency relation \cite{Creminelli:2012ed}
\be
\langle \zeta_{\vec q}\zeta_{\vec k_1}\zeta_{\vec k_2}\zeta_{\vec k_3} \rangle'_{\vec q \rightarrow 0} \sim \frac{1}{c_s^4}\mathcal{O}\left( \frac{q^2}{k^2}\right) P(q) P(k)^2 \;.
\ee
This is easy to understand since it receives contributions only from exchange diagrams: when $\partial^2\pi'$ corresponds to the external leg going to zero it will be trivially suppressed by $q^3$, when $\partial_i\pi$ corresponds to the external leg going to zero it will be contracted with both the other external leg which has some momentum $\vec{k}$ and the internal leg with momentum $-\vec{k}-\vec{q}$, which cancel at leading order in $q$. 

We checked also at the quartic level that the symmetry \eqref{eq:redsym} holds\footnote{For this check it is crucial to keep also the quadratic action and vary it with the last term of eq.~\eqref{eq:redsym}: this term cannot be neglected, because in going to conformal time it also gives an $H \epsilon'$ contribution.} and this will prevent any violation of the consistency relations. Violations are possible, like in the cubic case, if one considers quartic terms which do not respect the field redefinition symmetry.

\section{\label{sec:conclusions}Conclusions and outlook}
Given the simplicity of single-field inflation, it is certainly worthwhile exploring all the possible symmetries that can be imposed on its dynamics and their phenomenological consequences. Here we have studied the implications of imposing an approximate field redefinition symmetry $\phi \to \tilde\phi(\phi)$ on the inflaton. The predictions are very sharp since---after fixing the normalization of the spectrum---all correlation functions depend only on the speed of sound $c_s$ and are somewhat unusual, as a consequence of the slow decay of the decaying mode during inflation.

What we have studied represents another de Sitter limit of inflation, as inflation can (but need not) take place with the metric being exactly de Sitter. This parallels the case of ghost inflation \cite{ArkaniHamed:2003uz}, while another example has been studied in \cite{Cheung:2007st}. Like in the case of ghost inflation, the dynamics that may be responsible for modification of gravity in the late Universe, can be applied to inflation. This is not surprising, as models of modification of gravity often involve a scalar which defines a preferred foliation of space-time. And this is exactly what we need for inflation.

It is useful to think about this model as another corner of the EFT of inflation \cite{Cheung:2007st}. Starting from a general situation, the limit $\dot H \to 0$ kills the unitary gauge operator $g^{00}$, and therefore the standard spatial kinetic term of the inflaton. This is the limit of ghost inflation \cite{ArkaniHamed:2003uz}, when the spatial kinetic term is given by higher order spatial derivatives ($K^2$ and $K_{\mu\nu} K^{\mu\nu}$), while a standard time kinetic term $\dot\pi^2$ comes from the unitary gauge operator $(g^{00}+1)^2$. The symmetry that we discussed forbids any operator of the form $(g^{00}+1)^n$, so that also the time kinetic term is now given by the higher derivative operator $N^{-2} (\partial_i N)^2$. Of course these are only limiting cases: intermediate regimes in which various operators are relevant may have interesting features. We leave this to future investigations.

It is important to stress a relevant drawback of our model, i.e. its spatial non-locality: the Green function of $\pi$ shows instantaneous propagation of the signal as discussed in \cite{Blas:2011ni}. Most likely, this implies that our EFT cannot be embedded in a standard Lorentz invariant UV completion\footnote{We thank D.~Baumann for useful correspondence about this point.}. This is similar to what happens in models of $k$-inflation with superluminal speed of sound $c_s>1$.

\section*{Acknowledgments}
We thank D.~Blas, J.~Garriga, M.~Musso, O.~Pujolas, L.~Senatore, S.~Sibiryakov and Y.~Urakawa  for useful discussions. We are indebted to Austin Joyce for pointing out an error in the first version of the paper. J.N. is supported by FP7-IDEAS-Phys.LSS 240117. M.P. is supported by a MEC-FPU Spanish grant.

\appendix

\section{\label{app:after}Evolution after the field redefinition invariant phase}

In this Appendix we want to verify our intuitive arguments of Section \ref{sec:power}  in an explicit (toy) example. Let us add to the quadratic action \eqref{S2} a standard 2-derivative kinetic term\footnote{For simplicity we assume that the speed of sound of the kinetic term we added is the same as the one of the original terms.}
\beq
S = \int \mad^3x\,\mad\eta \left[\frac{M_\alpha^2}{2}(\partial\pi')^2 - \frac{M_\lambda^2}{2}(\partial^2\pi)^2 +\beta a^2 H^2 \left(\frac{M_\alpha^2}{2} \pi'^2-\frac{M_\lambda^2}{2}(\prt_i\pi)^2\right) \right] \label{S2_new} \;.
\eeq
We need $\beta \ll 1$ for the kinetic term discussed in this paper to dominate at Hubble crossing. In this case $\beta$ represents a small breaking of the field redefinition symmetry and the contribution of the kinetic term we added will become relevant when a mode is sufficiently long compared to the Hubble radius. What we want to check is that, up to corrections suppressed by $\beta$, $\pi$ remains constant during the out-of-Hubble evolution, until the mode becomes long enough to be dominated by the standard kinetic term. This will imply that the correlation functions calculated in the paper are actually the ones observed at late times. The equation of motion is given by
\beq
\prt^2\pi''-\beta H^2\frac{\mathrm{d}}{\mathrm{d}\eta} \left[a^2 \pi'\right] - \frac{M_\lambda^2}{M_\alpha^2} (\prt^4\pi-\beta a^2 H^2 \prt^2\pi)=0 \;.
\eeq
Out of the Hubble radius, i.e.~$(k/aH)^2 \ll 1$, there are three regimes of different evolution. For $\beta^{2/3} \ll  (k/aH)^2 \ll 1$, the terms proportional to $\beta$ are irrelevant and everything goes as discussed in the paper. The first term which becomes relevant is the Hubble friction and it is easy to realize that this is the only term one has to consider in addition to the original Lagrangian in the window $\beta \ll (k/aH)^2 \ll \beta^{2/3}$. Finally, in the regime $(k/aH)^2 \ll \beta$, only the terms proportional to $\beta$ are relevant and $\pi$ behaves as in standard inflation. It is simple to follow the evolution from one phase to the other in the long wavelength limit. First of all notice that $\pi = const$ is a good solution in any phase and in the transition regions for a mode which is well outside the Hubble radius, i.e. in the $k \to 0 $ limit. This can be seen explicitly in the equation and follows from the general conservation of $\zeta$ on super horizon scales (which within our approximations implies the conservation of $\pi$ as $\zeta = -H \pi$, with constant $H$). Moreover, the velocity becomes irrelevant, $\dot \pi \ll H \pi$, before the terms proportional to $\beta$ start playing any role, and this implies that $\dot\pi$ can be neglected when matching to the next phase. There is no mode mixing and $\pi$ remains constant all along. It is easy to check this behaviour numerically.

The same reasoning works if we allow $\beta$ to be time-dependent, i.e.~dependent on the background value of $\phi_0 = t$. This describes the fact that the field redefinition symmetry will be badly broken at the end of inflation and $\beta$ will become large. It is straightforward to check that also in this case $\pi = const$ is a good solution so that, for modes well ouside the Hubble radius, i.e.~$\dot\pi \ll H\pi$,
the field remains constant while the symmetry gets broken. Notice that the logic is exactly the same one uses in the case of standard inflation to justify the conservation of $\zeta$ through the unknown reheating phase. As in that case we expect the same arguments to be valid non-linearly in the amplitude of $\zeta$, so that each $n$-point function remains the same when out of the horizon.

\section{\label{app:constraints}Constraints and the validity of the decoupling limit}

In the main text we have calculated everything in terms of $\pi$, focussing on the decoupling limit, (i.e.~neglecting its effects on the metric) and then converting the results in terms of $\zeta$. The logic behind it is that we expect the corrections coming from the effect of $\pi$ on the metric to be subleading in $1/M_P^2$, and therefore negligible when $M_\lambda^2 \ll M_P^2$ and $M_\alpha^2 \ll M_P^2$. However the model we are describing is sufficiently unconventional to warrant a check of this intuitive explanation. Let us calculate the power spectrum of $\zeta$ directly in the $\zeta$-gauge, i.e.~setting to zero the $\pi$ perturbations. 

Starting from the action \eqref{S_uni}, we go through the standard procedure \cite{Maldacena:2002vr} of solving the constraint equations and plug the solution back into the action.
We use the ADM splitting of the metric \eqref{adm}.
Defining $N=1+\delta N$ and $N^i=N^i_T+\partial_i \psi$, with $\prt_iN^i_T=0$, the linearized constraint equations obtained by varying with respect to $N$ and $N^i$ respectively are given by
\begin{equation}
\left (1+\frac{3}{2}\frac{M_\lambda^2}{M_P^2}\right)\left(\partial^2\psi-3(\dot\zeta-\delta N H)\right)+\partial^2\left(\frac{\zeta}{a^2H}\right)+\frac{M_\alpha^2}{M_P^2}\frac{\partial^2\delta N}{2a^2H} = 0 \nonumber 
\end{equation}
\begin{equation}
\partial_i\left [(\delta NH-\dot\zeta)\left (1+\frac{3}{2}\frac{M_\lambda^2}{M_P^2}\right)+\frac{M_\lambda^2}{2M_P^2}\partial^2\psi\right]=0 \;.\label{conseq}
\end{equation}
We can now solve these equations at first order in $M_\alpha^2/M_P^2$ and $M_\lambda^2/M_P^2$ and plug the solutions back into the action. After some work, we obtain   
\begin{align}
S_\zeta &=  \int \mad^3x\,\mad\eta\left(\frac{M_\alpha^2}{2 H^2}(\partial\zeta')^2 - \frac{M_\lambda^2 }{2 H^2}(\partial^2\zeta)^2 \right)\;, \label{Szeta2}
\end{align} 
which is the action given in \eqref{S2} with $\pi=-\zeta /H$ as expected. The action above will contain additional terms suppressed by powers of $M_\alpha^2/M_P^2$ and $M_\lambda^2/M_P^2$.

\end{document}